\documentclass[review]{elsarticle}

\usepackage{lineno,hyperref}
\modulolinenumbers[5]

\usepackage{empheq}

\journal{Nucl. Instr. Meth. Phys. Res.}

\begin{document}
\begin{frontmatter}
\title{Phase I results with the Large Angle Beamstrahlung Monitor (LABM) with SuperKEKB Beams}

\author[tabuk]{ R. Ayad}

\author[wayne]{G. Bonvicini}
\author[wayne]{S. Di Carlo}

\author[cinvestav]{E. De La Cruz-Burelo}

\author[wayne]{H. Farhat}
\author[kek]{J. Flanagan}
\author[wayne]{S. Gamez Izaguirre}
\author[wayne]{R. Gillard}
\author[cinvestav]{Michel Enrique Hern\'andez-Villanueva}

\author[kek]{H. Ikeda}

\author[kek]{K. Kanazawa}

\author[tabuk]{B.O. Elbashir}

\author[sinaloa]{P.L.M. Podesta-Lerma}
\author[sinaloa]{D. Rodr\'iguez-P\'erez}

\author[Puebla]{G. Tejeda Mu\~noz}

\address[tabuk]{Department of Physics, Faculty of Science, University of Tabuk, Tabuk 71451}
\address[wayne]{Wayne State University, Detroit, Michigan 48202}
\address[kek]{High Energy Accelerator Research Organization (KEK), Tsukuba 305-0801}
\address[sinaloa]{Universidad Autonoma de Sinaloa, Culiac\'an, Mexico}
\address[Puebla]{Benemerita Universidad Autonoma de Puebla, Puebla, Mexico}
\address[cinvestav]{Centro de Investigacion y de Estudios Avanzados del IPN, Mexico City, Mexico}

\begin{abstract}
We report on the SuperKEKB Phase I operations of the Large Angle Beamstrahlung Monitor (LABM). The detector is described and its performance characterized using the synchrotron radiation backgrounds from the last Beam Line magnets. The
backgrounds are also used to determine the expected position of the Interaction Point (IP), and the expected background rates during Phase II.
\end{abstract}

\begin{keyword}
\texttt{Beamstrahlung}\sep \texttt{luminosity}\sep \texttt{Beam-monitoring}\sep 
\end{keyword}

\end{frontmatter}
%\linenumbers
\section{Introduction}
The Large Angle Beamstrahlung Monitor (LABM) is a direct monitor of the beam
parameters at the Interaction Point (IP). Built by a collaboration of five
institutions headed by WSU, the LABM consists of four narrow angle telescopes
which collect light from the IP, divide it into two polarizations and four
wavelengths, and count photons by means of 32 total photomultipliers (PMT). The main background separation
between beamstrahlung signal and synchrotron radiation backgrounds exploits the very forward angular pattern of radiation from a magnet, whereas the radiation from the very short beam-beam region is much broader in angle~\cite{bonvicbn}.

The four telescopes account for observation of the emitted light for both beams
both above (azimuthal position of 90 degrees, or up) and below (270 degrees, or down) the
Beam Pipe (BP). The redundancy improves the measurement of a number of systematics,
including possible misalignments of the beam trajectory with the detector
axis.

\begin{figure}[ht]
    \centering
    \includegraphics[width=0.8\textwidth]{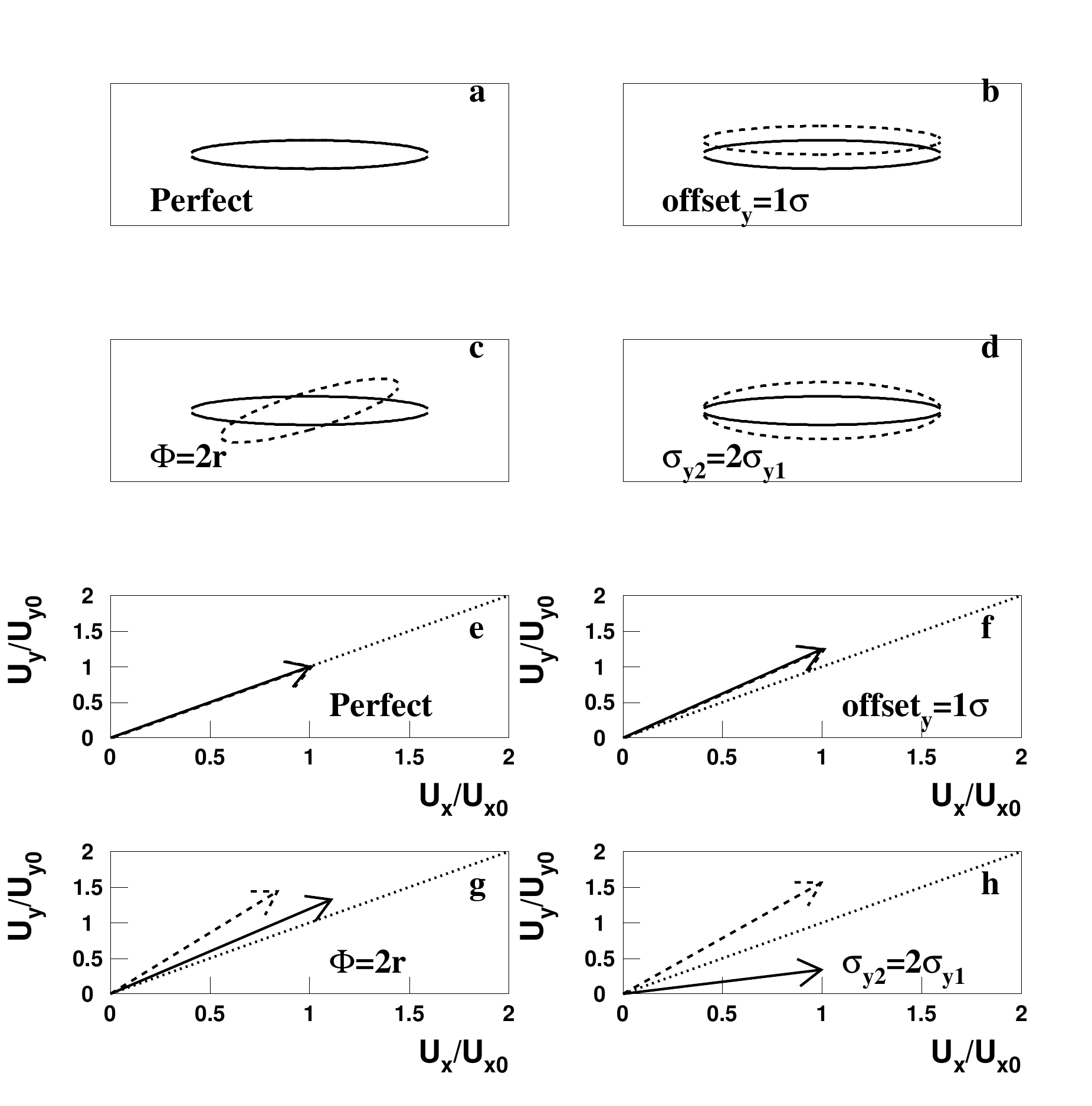}
    \caption{Graphic representation of how large beamstrahlung variables
      relate to beam-beam collisions. The beams
      are represented as ellipses in the transverse plane
      at the moment of collision.
      Solid arrow corresponds to solid beam
      and dashed arrow to dashed beam. The arrows represent the observed
      polarized rates $U_x, U_y$ and produce a different pattern for a different
      beam-beam mismatch.}
    \label{fig:arrows}
\end{figure}

The detector works by building observables from the ratio of polarizations,
and in particular it determines directly at the IP the relative height of the
two beams and their relative vertical offset. This is described graphically
through the use of beamstrahlung diagrams, Fig.~\ref{fig:arrows}. Direct
IP observations are rare, and the direct determination of the beam heights
at the IP particularly crucial given the luminosity limitations
encountered at KEKB.

The detector was first installed in Summer 2015, then installed a second time
during Fall 2015, and after some upgrades, was operational in February 2016,
a few days before beams started circulating. We have taken data continuously
during Phase I. 

The paper describes how we were able to
successfully see light coming from the IP, shows that the detector works
as specified, and describes the small modifications done to the LABM to
overcome the problems seen with two of the four telescopes. The observed
background rates are used to perform side-to-side (positron-to-electron)
cross checks and to determine the expected background rates during Phase II.

\section{Detector description.}

To accurately study light emitted from the IP, light needs to be extracted from the Beam Pipe, transported to a relatively low radiation area, and its spectral and polarization components counted separately. 

Each telescope has four mirrors at 45 degrees, at locations between 451 and 477cm from the IP, which reflect incoming light into a Fused Silica Vacuum Window located opposite in Beam Pipe azimuth, see Table~\ref{tab:geometry}. Once outside the Beam Pipe, light is transported over a length of 8-10 meters through four Optical Channels, consisting of several dark Aluminum pipes and 45 degrees mirrors, to a sheltered area where light is analyzed inside two Optics Boxes, one for the electron ( or ``Oho'' side of the detector)
and one for the positron beam ( or ``Nikko'' side of the detector).
Each Optics Box has two optical tables, servicing both viewports on that side. Inside the Optics Boxes, light is separated into components and counted.

\begin{table}%{htb}
\centering
  \caption{Optical and geometrical parameters of the LABM. The $\phi$ acceptance
  is the same for all detectors and equals $1/(40\pi)$, or 0.0080.}
\label{tab:geometry}
\begin{tabular}[t]{c|c|c|c|c}
\hline
 Telescope & distance IP(m) & Spot size at Box (mm) & $\theta_{min}$ & $\theta_{max}$\\
 \hline
Oho Down & 4.51 & 6.52 & 8.43 & 8.87 \\
Oho up & 4.57 & 6.00 & 8.32 & 8.76 \\
Nikko Down & 4.77 & 6.60 & 7.97 & 8.39 \\
Nikko Up & 4.70 & 6.11 & 8.08 & 8.50 \\
\hline
\end{tabular}
\end{table}

The first component along the line are the vacuum mirrors, made out of Beryllium and having sizes of 2$\times 2.8$mm$^2$, which corresponds to a square angular acceptance once the mirror inclination is taken into account. The mirrors are
passively cooled, with the copper base carrying heat out to copper fingers which are themselves air cooled by small fans. The mirrors are small to minimize RF heating, and their inner edge is 38 mm from the nominal Beam Line.

The small mirrors represent the first of two collimators in a simple two-collimator optics to limit the telescope angula acceptance and reject collinear backgrounds. Given that the detector is sensitive in the $350<\lambda <650$~nm range, diffraction effects are of the order 0.18-0.33 mrad, which are small compared to the about 8~mrad angles of observation for the signal~\ref{tab:geometry}.

On exiting the BP, light enters the Optical Channel. This is a system
of aluminum pipes and 45 degrees mirrors (``Elbows'') to transport light
to the Optics Boxes, located in a radiation sheltered area.
All mirrors are Surface First UV Enhanced Aluminum Elliptical Mirrors from Edmunds Optics for 45 degrees reflection. The first mirror in each OC is one inch minor axis, the second is 2 inches minor axis, and all subsequent mirrors in every OC are 4 inches minor axis. The first two in each line, called primary and secondary, are remotely controlled for pointing purposes, through stepper motors and their controllers, themselves activated by our Slow Control Software. The aluminum pipes likewise are 1.5 inches inner diameter for the connector between BP and primary mirror, 2 inches for the connector between primary and secondary mirrors, and 4 inches for all subsequent pipes. All OC have black paint in the first two pipes and two to four 1 cm optical baffles at points along the OC to reduce reflections.

Each OC is of different length with different turns, as the OC have to dodge numerous pieces of equipment. Over a length of about 10 meters, light is transported through a 1.5 meters concrete floor to the Optics Boxes.

Entering the Optics Box the light beam (with a typical divergence of 2.2 mrad)
encounters a broad band Wollaston prism from Lambrecht Corporation
which splits the beam into two beams,
with local transverse polarization $x-$ and $y-$. The two polarized beams are
separated by 20 degrees. Each beam is then guided by two mirrors onto a ruled
grating, a type of grating which transmits maximal intensity to the first
order peak, from Thorlabs.
The rainbow exiting the grating is guided onto 4 PMTs, each
observing approximately 75 nm bandwidth, where photons are counted.
For Phase I only, gratings with a narrower angular spread were used, to have a
``blind'' PMT, not illuminated by the incoming beam, for the purpose of
measuring Cherenkov backgrounds in the PMT window during data taking.
These backgrounds were found to be vanishingly small.

\begin{figure}[ht]
    \centering
    \includegraphics[width=0.8\textwidth]{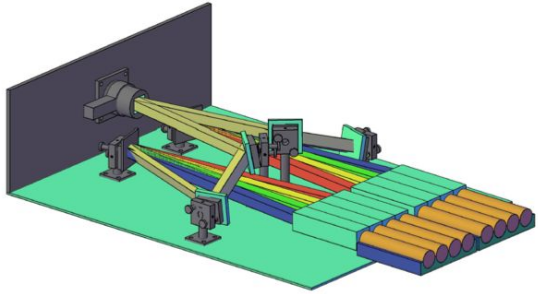}
    \caption{Schematics of one half of an Optics Box, which views one viewport.
      The light beam enters the Box, and is split into two polarized beams by
      a Wollaston prism. Each beam is then spread by a grating and a four
      component spectrum is measured by quartets of photomultipliers.}
    \label{fig:obschematics}
\end{figure}

Fig.~\ref{fig:obschematics} shows an Optics Table, which is
half an Optics Box schematics.
The PMTs are Hamamatsu R6095, and they are mounted on a remotely controllable
conveyor belt. They can be switched onto different viewports, and in particular
PMTs observing the up telescope can be rotated to observe the down telescope
for the purpose of measuring out some of the systematic errors in the determination of beamtrahlung asymmetries which provide information about beam-beam parameters.
Numerous pictures of our installation can be found
at the website~\cite{gallery}.

\section{Electronics and DAQ.}

The PMTs signals are amplified, discriminated by a -30mV threshold, and sent
through cables to our counting room (which also hosts the electronics of
other beam monitors), located respectively 25 and 80 meters
(in cable length) away from the two counting boxes. There a CAEN scaler V830
counts data at (currently) 1Hz rate, although our DAQ permits data taking at
frequencies far exceeding relevant seismic frequencies
to study seismic effects on the beam alignment. A PC donated by KEK collects
and writes all data, which include the 32 PMT and various other beam monitors
around the ring, obtained through the EPICS network. The PC is accessible
through ftp or other Internet protocols, and we are therefore fully
independent in analyzing our data as well as monitor correlations. For
the analysis presented here, the only SuperKEKB information used were the
instantaneous beam currents.

\section{Calibration and alignment.}

At time of assembly, we place a diode in front of the vacuum window and
assemble the Optical Channels starting from the beam Pipe. As the Channel is
assembled, we can observe visually the reflected image of the diode by removing the
mirror and back side of the next elbow. This trick allows us to look inside
the OC after a new pipe and elbow are assembled.

If the image is not centered,
the elbow under study is re-oriented until the image is centered. The process
is repeated down the Channel. The last elbow entering the Optics Box can not
be aligned this way. However, the last elbow is located about 8 cm in front
of the Optics Box entrance, and the previous leg of the Channel is 550 to 570
cm depending on the OC.
No significant beam misalignments can occur with such a geometry.

PMT calibration by using the conveyor belt and light from a diode
was first done successfully in
Summer 2015. The calibration is done by illuminating the OB
with a fixed light source placed just in front of the vacuum window, while
rotating the 16 PMTs on, say, the positron side through the 16 viewports
of that side. 192 measurements are obtained for 32 degrees of freedom,
allowing the determination of the relative efficiency of
the 16 PMTs observing the positron side. The procedure was also done on the
electron
side. In the future, swapping some PMTs from one side to the other side should
allow a relative efficiency determination for all 32 PMTs.

\section{Phase I data.}

The goals of Phase I data taking were as follows: first, generally commission
the device, second, prove that we could point the telescopes (with an acceptance
of approximately 1mrad$^2$) to the IP, and third, cross check that the two
sides were seeing rates consistent with one another. As only one beam at a time
was present in the Beam Pipe, all analysis was done with synchrotron radiation
from the last magnets in the Beam line. Tables ~\ref{tab:angles} 
and ~\ref{tab:dipoles} provide the geometrical parameters of detector and last
magnets.

\section{Bending magnets}
\begin{figure}[ht]
    \centering
    \includegraphics[width=0.8\textwidth]{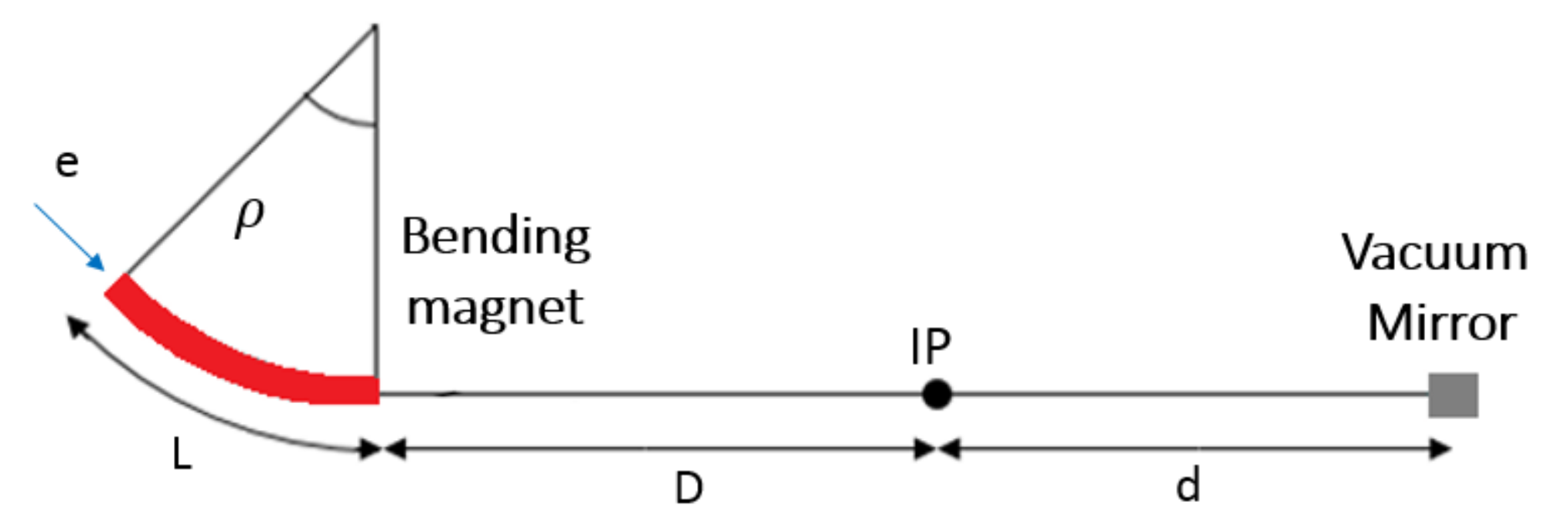}
    \caption{Schematic to find the length of the sweep from the geometry of the interaction region.}
    \label{fig:bendingMagnet}
\end{figure}

During Phase I, we measured the synchrotron radiation (SR) emitted when the beams are bent while traveling inside the bending magnets (BM) near the IP. It is
convenient to discuss the properties of such SR sweeps before looking at
the angular scans.
The BMs are located on each side of the IP, and their position respect to the LABM vacuum mirrors is shown in Fig.\ref{fig:bendingMagnet}. Since the SR is emitted while the beam progresses through the BM, the image produced is a sweep. The length of the sweep depends on the length L and radius of curvature $\rho$ of the BM. With the help of Fig.\ref{fig:bendingMagnet}, the angle $\theta_{s}$ subtended by the sweep, as seen from the LABM, is calculated as:

\begin{equation} \label{eq:sweepLength}
%\theta_{s} = \arctan \Bigg( \frac{\rho}{d+D} [1 - \cos(L/\rho)] \Bigg)
\theta_{s} \approx \frac{L^{2}}{2\rho(D+d)}
\end{equation}

where D is the distance from the edge of the bending magnet to the IP, and d the horizontal distance from the vacuum mirror to the IP.

The angle subtended by the SR sweep is tabulated, for each of the 4 LABM channels, in Table \ref{tab:angles}. We have a sweep of 1.5 mrad for the Nikko side, and a sweep of 0.8 mrad for the Oho side. We define the spacial resolution of the LABM as the diameter of the spot from which the LABM accepts radiation. For phase I, the angle subtended by this diameter was 1.4 mrad. Therefore, our angular resolution is smaller than the Nikko sweep and larger than the Oho sweep. Therefore, we expect that we will be able to measure the Nikko sweep as an image of length 1.4+1.5+1.4=4.3 mrad, and the Oho sweep as an image of length 1.4 mrad.

\begin{figure}[ht]
    \centering
    \includegraphics[width=0.8\textwidth]{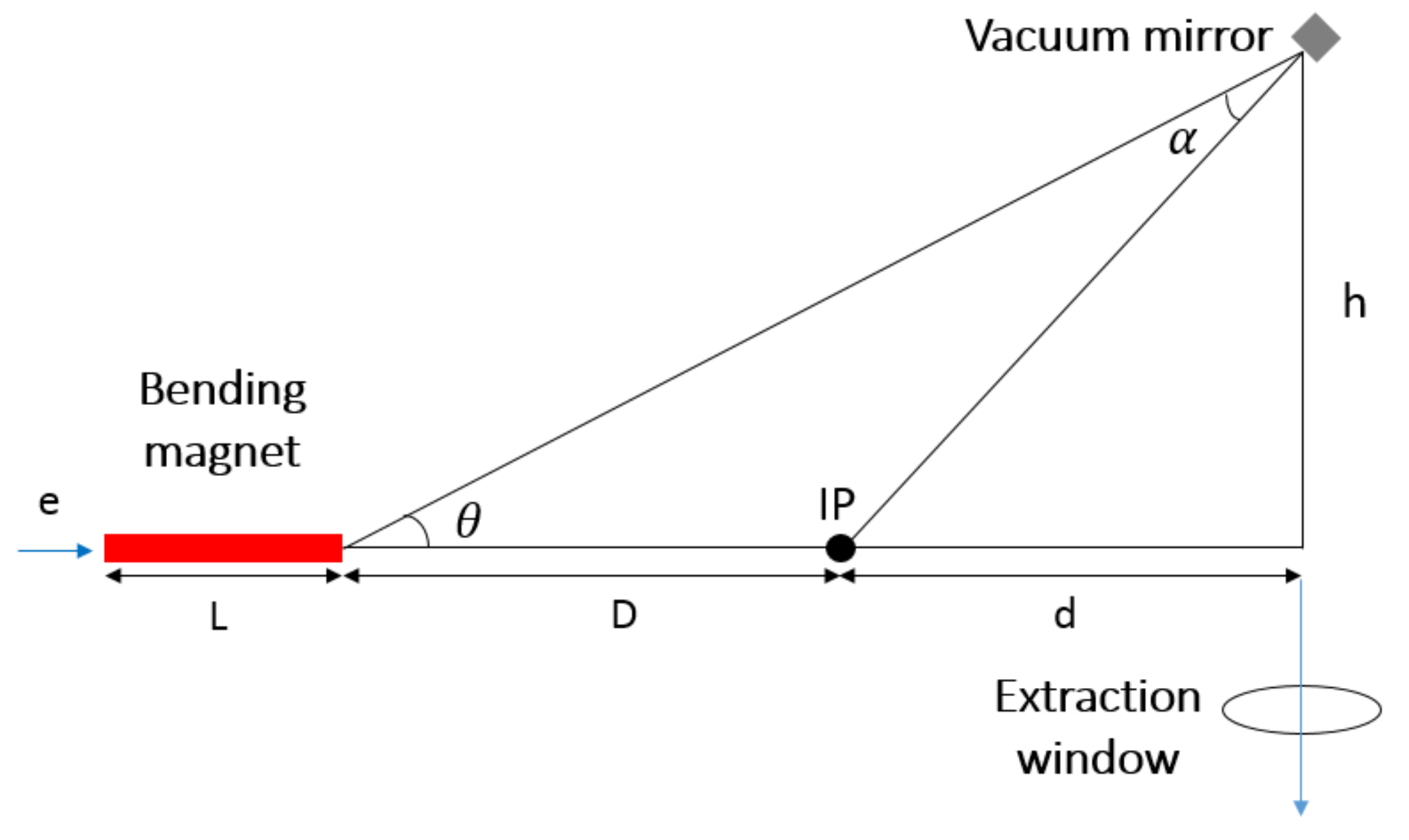}
    \caption{Schematic to find the angle $\alpha$ that separates the sweep from the IP. The IP will be located at an angle $\alpha$ below the position of the SR sweep.}
    \label{fig:angleIP}
\end{figure}

Scanning the solid angle inside the beam pipe, we can find the angular position of the SR sweep, and then calculate the angular position of the IP. At the exit of the bending magnet, the SR collected by the LABM is emitted at an angle $\theta$ from the axis of the beam pipe, as shown in Fig.\ref{fig:angleIP}.

The angular separation of the IP from the SR sweep can be calculated as

\begin{equation} \label{eq:IPangle}
\alpha = \arcsin \Bigg( \frac{D}{d}  \frac{\sin \theta}{\sqrt{1+\frac{h^{2}}{d^{2}}}} \Bigg)
\end{equation}

where h is height of the vacuum mirror from the beam pipe axis. To find the position of the IP from the image of the scan, we need to move by an angle $\alpha$ in direction -Y. Analogously, for an ''up'' channel, we would need to move by an angle $\alpha$ in direction +Y. All the relevant quantities described in this section are shown in Table~\ref{tab:angles}, with the magnets lengths and radii
of curvature shown in Table~\ref{tab:dipoles}.

\begin{table}[ht]
    \begin{tabular}{ l | l | l | l | l  | l | l }
    \hline
Channel & d~(m) & $D$~(m) & h (m) & $\theta \, (mrad)$ & $\alpha \, (mrad)$ & $\theta_{s} (mrad)$\\ 
\hline
Oho Down & 4.51 & 8.02 & 0.04 & 3.19 & 5.68  & 0.80\\ 
\hline
Oho Up & 4.57 & 8.02 & 0.04 & 3.18 & 5.57 & 0.80\\ 
\hline
Nikko Down & 4.77 & 5.11 & 0.04 & 4.05 & 4.34 & 1.53\\ 
\hline
Nikko Up & 4.70 & 5.11 & 0.04 & 4.08 & 4.43 & 1.54\\ 
    \hline
    \end{tabular}

    \caption{The relevant positions and angles of LABM and BM discussed in this section. $d$ is distance to the IP.}
    \label{tab:angles}
\end{table}

\begin{table}%{htb}
\centering
\caption{Last magnet parameters before LABM in each Beam Line. $\rho$ is the
  curvature radius and $d_{LABM},\theta_{LABM}$ are the distances and angles of the
magnet end to the LABM vacuum mirrors.}
\label{tab:dipoles}
%\begin{center}
\begin{tabular}[t]{c|c|c|c|c|c}
\hline
 Side & $\rho$(m) & Length (m) & $d_{LABM}$ (m) & $\theta_{LABM}$ (mrad) & Angle $\gamma$\\
 \hline
 LER & 148.686 & 2.119 & 9.37 & 4.2 & 33 \\
 HER & 609.065 & 3.49 & 12.11 & 3.2 & 44 \\
\hline
\end{tabular}
\end{table}

\section{\bf Data analysis.}

In the following, all data are pedestal subtracted rates.
As stated in Section~2, the telescopes have a double collimator optics. The
first collimator is the small vacuum mirror, and the second is a sheet metal
collimator placed directly before the 19 mm square Wollaston prism.

\begin{figure}[ht]
    \centering
    \includegraphics[width=0.8\textwidth]{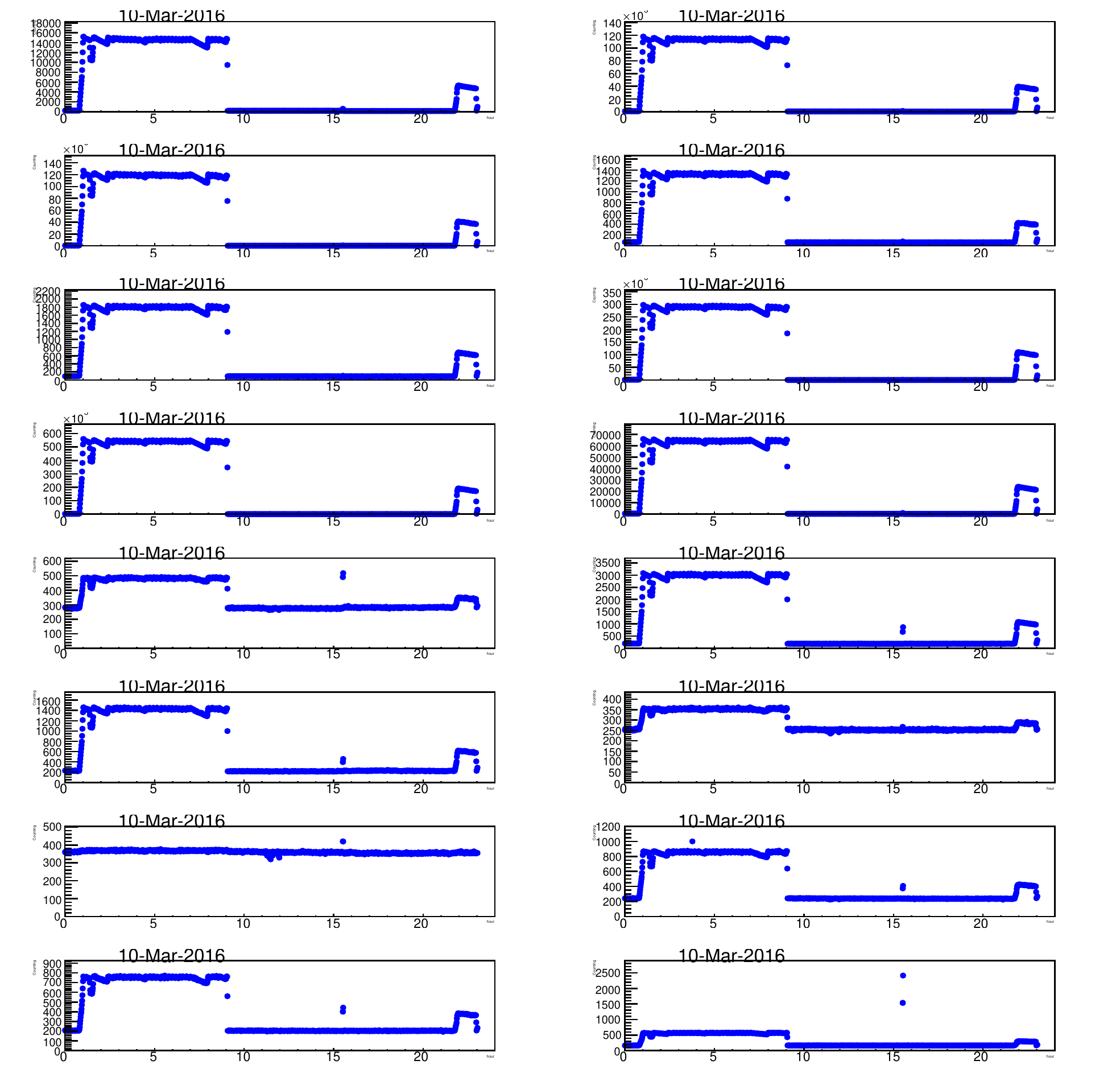}
    \caption{Oho PMT Rates versus time for March 10, 2016.The last fill with the 12mm collimator and first fill with the 8mm collimator are included in the time window. We notice that the rate, for the same current of the radiating beam, drop to about 1/3 of the initial value. Time (hour) on the horizontal axis. Rate (Hz) on the vertical axis.}
    \label{fig:coll}
\end{figure}

We started taking data with a 19 mm collimator (no collimator at all, with the acceptance being limited by the size of the Wollaston prism). As we understood the images, we moved to 15 mm collimators, then 12mm, then 8mm. 7mm square collimators were installed after Phase I, but 2mm collimators will be the final configuration. Fig.~\ref{fig:coll} shows at a glance the rate of all the Oho PMTs over one day, with 60 mA beams before and after an access during which the collimators were changed from 12mm to 8mm. One can see that the detector is reasonably dark, quiet, and that the measured rates follow from the solid angle reduction by the collimator.

\begin{figure}[ht]
    \centering
    \includegraphics[width=0.8\textwidth]{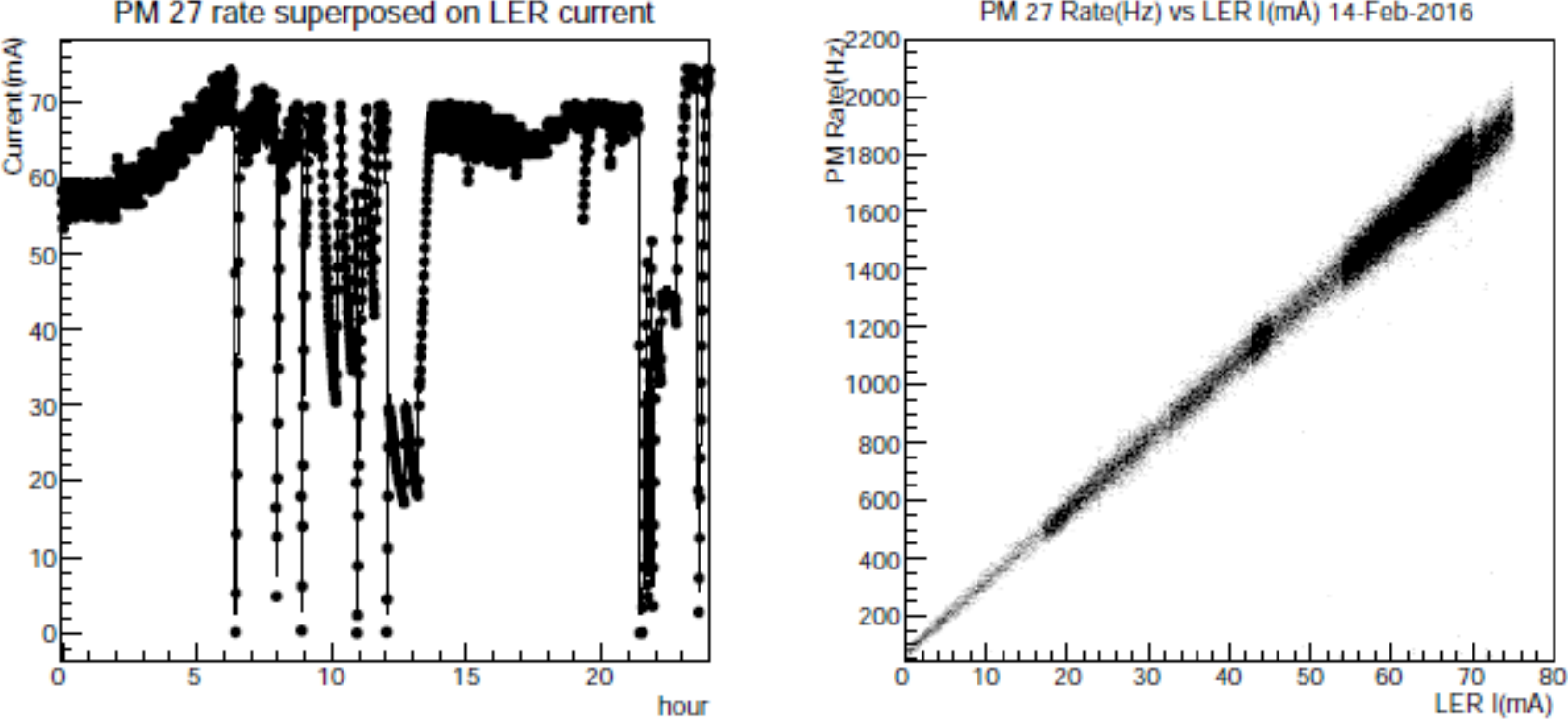}
    \caption{Left: Rate-current correlation for one of the LABM PMTs. Rate (solid line) and current (marker) superimposed on the left. Time (hour) on the horizontal axis. Rate (Hz) on the vertical axis. Right: Rate versus current. Current (A) on the horizontal axis. Rate (Hz) on the vertical axis. We notice that the signal from the PMTs is linearly correlated to the current of the radiating beam.}
    \label{fig:linearity}
\end{figure}

\begin{figure}[ht]
    \centering
    \includegraphics[width=0.8\textwidth]{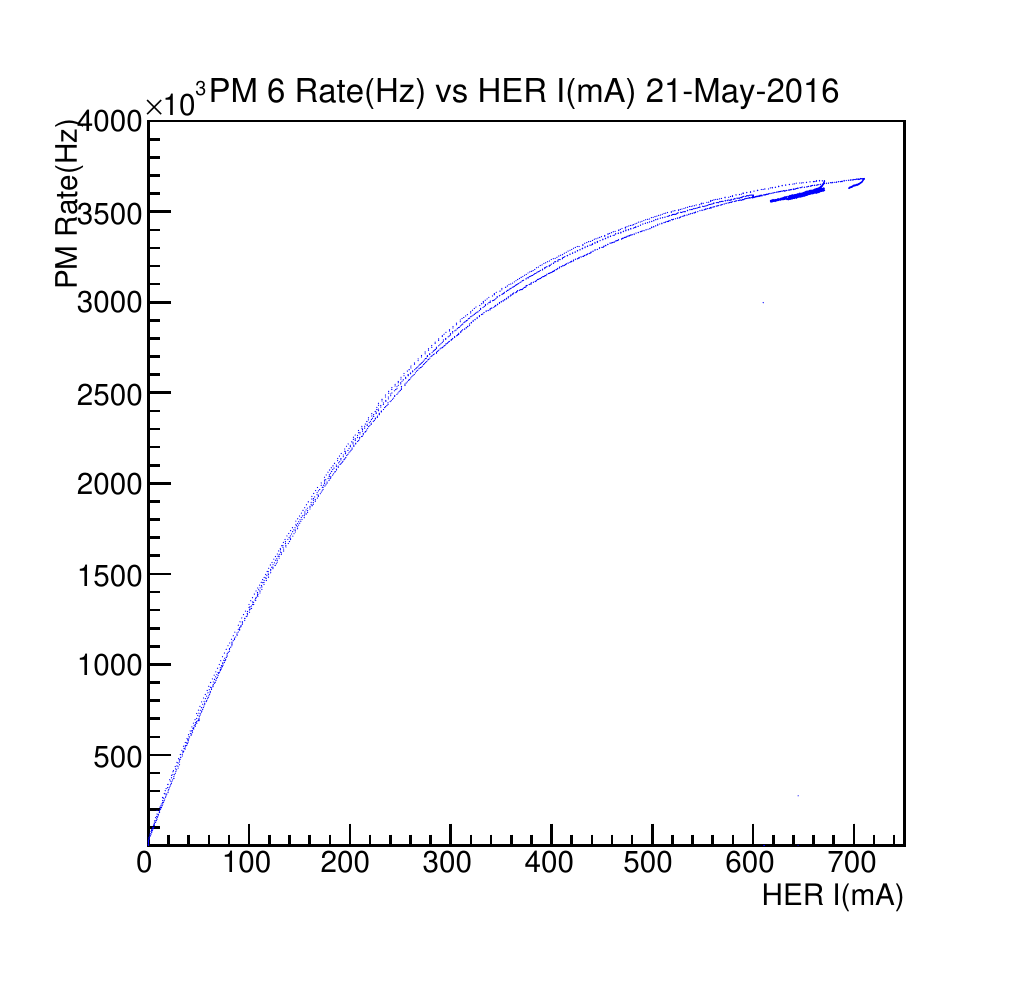}
    \caption{Rate-current correlation for one of the LABM PMTs, taken
      on a high current day. The saturation is due to discriminator dead time.}
    \label{fig:cor}
\end{figure}

We have very often taken data during widely varying beam conditions. The condition for significant data taking is that the rates seen be proportional to the beam current. This is certainly the case for low beam currents, Fig.~\ref{fig:linearity}. After our main tests performed at currents of order 100 mA, we parked the detectors at the predicted IP angle. As the beam currents increased, saturation effects started to become apparent, Fig.~\ref{fig:cor}. 
The curve fits best with a 6.0MHz saturation rate which
is in rough agreement with a (predicted) saturation rate of 7.5MHz. In any case
each PMT response curve was fitted with its own dead time curve, and we extract
the true rates according to

\begin{equation} \label{eq:correction}
R_{true}=\frac{R_{obs}R_{max}}{R_{max}-R_{obs}},
\end{equation}
where $R_{max}$ is the measured saturation rate.

Using the specific rate (Hz/mA) allowed us to take long scans to study details of the light reaching our detector. With the detector stable and operating properly the crucial question is if we can deduce the location of the Interaction Point (IP), based on the observed light pattern. If the location can be determined, then direct observation is possible. It is noted that eventually, at nominal conditions, the IP will be by far the brightest spot in our images. However the beamstrahlung signal scales roughly like $I_+I_-^2$ for the electron beam, and also like $1/\sigma_x^2$, where $\sigma_x$ is the transverse horizontal width, common to both beams. If collision data are to be taken, the backgrounds will have to be substracted. Besides finding the IP based on the observed pattern, we measure the expected background rates for that point.

The detector response was measured in November 2015 with a long series of runs using a white light source and a point-like white reflector placed just above
(or below) the vacuum windows. The measurements focused on the relative calibration of the PMTs, which are known to vary by 30\%, whereas each piece in the optical lines had been measured on the bench at WSU with accuracies well below 1\%.
 
The source spectrum is constant, and the source is assumed to be unpolarized to high precision. However the source, which consists of a lamp external to the Optical Channel and a small, white diffuser placed directly above or below the vacuum window, needs to be re-positioned from the Up to the Down detector when calibrating the detector, so that different rates can be observed.

Subsequently we take data while rotating the PMTs by means of the conveyor belts in the Optics Boxes, in 12 different configurations, both the PMTs relative efficiencies and the relative efficiencies of the optical efficiencies of each channel can be calculated (this corresponds to a fit with 112 data and 24 free parameters). These corrections are convoluted in the analysis presented below.

With increasing currents we observed a saturation effect due to the
discriminators dead time, Fig.~\ref{fig:cor}. With the current electronics, we
will run with smaller collimators and light attenuators during Phase II.

\section{Properties of LABM angular scans.}

Angular scans were performed to try and identify the IP. By orienting either the primary or the secondary mirror using remotely controlled motors, we were able to accept light coming from different angles inside the Beam Pipe. Since the primary and secondary mirrors are very close (typically 10 to 20 cm out of optical lines exceeding 15 meters from IP to the Optics Box), we have tested numerous times that we could obtain identical images by scanning with only one of the two mirrors, while keeping the other fixed. The angle to step motor conversion factors are all published in our Belle II Twiki page~\cite{labmpage}, as well as which coordinate and which sign for each motor.

We have chosen reference frames where the $+y$ axis always points up and the $+x$ axis is always pointing towards the center of the ring. As a result, two of our angular frames are right-handed and two are left-handed. The scans are done on a 26X41 grid of points, typically spanning 20X20 mrad$^2$. Each point and each current are taken five times and averaged before moving the telescope, and the specific rates are plotted. Initially, scans were taken with 19 mm collimators (the Wollaston prisms themselves). As we found the luminous areas, we restricted the angular range of the scans and decreased the size of the collimators.

\begin{figure}[ht]
    \centering
    \includegraphics[width=0.8\textwidth]{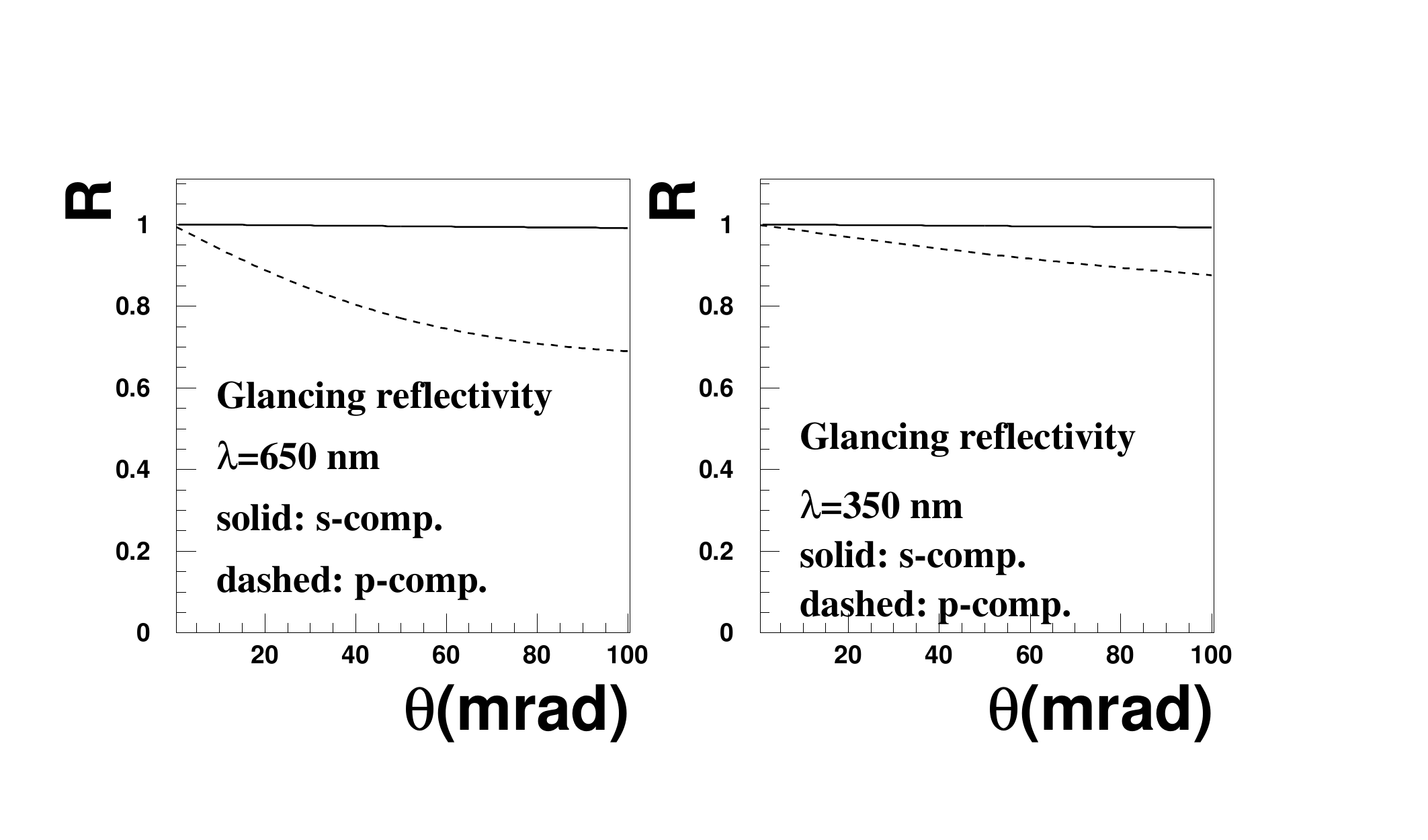}
    \caption{TiN glancing reflectivity for both s- and p- components
as a function of glancing angle. First plot: $\lambda=650$ nm. Second plot:
 $\lambda=350$ nm.}
    \label{fig:tin}
\end{figure}

Numerous images are always present in our scans, due to reflections inside the Beam Pipes. While the HER pipe is brushed aluminum, highly reflecting at low angles and for all (visible) wavelengths considered, the LER pipe is coated with sputtered titanium nitride(TiN), whose optical characteristics are not well known. We contacted the authors of Ref.~\cite{materials}, and with their help we were able to calculate the low angle reflectivity of TiN, presented in Fig.~\ref{fig:tin} for our lowest and highest wavelengths. We can not automatically assume that the brightest image be the directly observed beam.

Reflections can come from light emitted at lower angles than the actual angle of direct observation. Also, reflections at, for example, 45 degrees in azimuth will not produce any significant $x-y$ polarization since the reflection polarizes light at 135 degrees, with equal components along the vertical and horizontal axes.

% xxx
The method used to identify the IP in our scans goes as follows:

\begin{itemize}
\item complete a scan. Look for the main sweeps or spots to identify the incoming beam, radiating from the last bending dipole, whose location and strength is known (see Table~\ref{tab:dipoles}). The last spot in the sweep (or more precisely, the outermost point of the sweep in the horizontal plane) is where the beam exits the dipole.
\item  given the known geometry of the IR, the IP will be a known number of milliradians above or below the last spot.
\item reflected radiation is not necessarily redder and more polarized than the direct image, because reflections can happen near the 45 degrees area of the Beam Pipe, and because reflections are typically from radiation emitted at a lower angle than the direct observation. We check that the polarization and spectral pattern,
\end{itemize}

Our detector has a triangle shaped angular acceptance, with base approximately 1 mrad with current collimators. Both sweeps are fully visible from our vacuum mirrors. The LER sweep will be approximately 3.3 mrad in angular width, the HER will be 1.4 mrad. As saturation effects decrease the quality of a picture, images were taken with PMTs that presented the least saturation.

\section{Angular scans.}

\subsection{Nikko Down}

\begin{figure}[ht]
\centering
\includegraphics[width=0.8\textwidth]{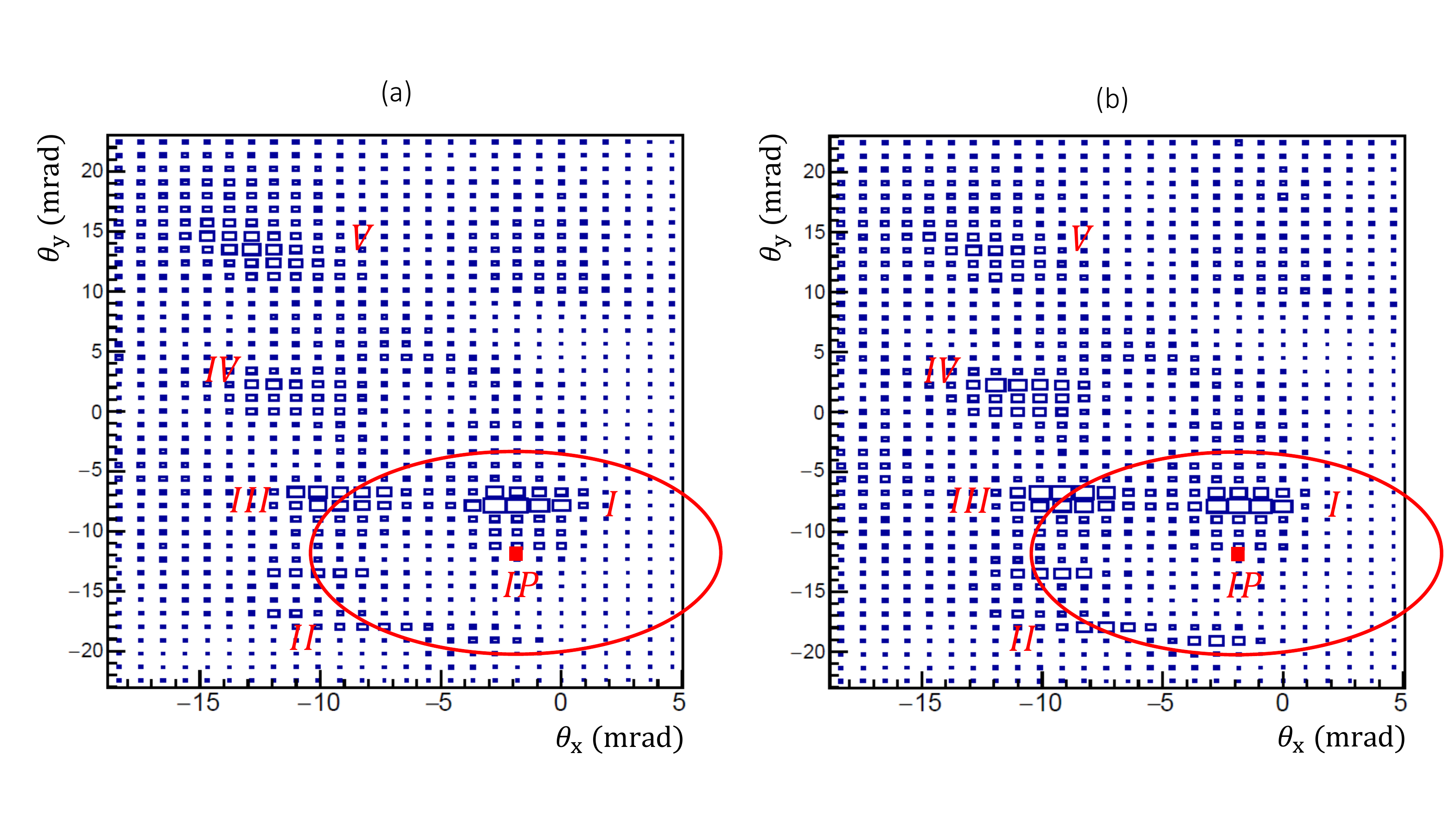}
\caption{Nikko Down angular scan with ``red'' PMT. First plot: x-polarization.
  Second plot: y-polarization. The IP is marked as a
  red square, the structures marked I-IV are discussed in the text. The
  waist of the Beam Pipe at the IP is shown as a red ellipse. }
\label{fig:nikkodown}

\end{figure}

The Nikko side (observing the LER, or positrons) faces a hard last bend (Table~\ref{tab:dipoles}). The angular scan in the red x- and y-PMTs are shown in Fig.~\ref{fig:nikkodown}.

The single sweep on the left (reflection I) has characteristics consistent with being the direct image.

\begin{itemize}
\item It has the right angular size, after detector effects are taken into account,
\item as the beam progresses through the sweep, the angle $\theta$ to the vacuum mirror decreases. We observe a  decrease of 60\% total intensity from left to right. 
\item as mentioned before, the beam sweep, projected to the front of the magnet, is 3cm long. This means that the azimuthal angle to the vacuum mirror changes from about -53 degrees to about -90 degrees. Based on the azimuth, we expect a substantial polarization change across the sweep. This is best done with the ``blue'' PMT, since the onset of the short magnet regime is faster with higher frequencies. In the four main pixels of the sweep the polarization $R_x/R_y$ goes from 9.3 to 19.4 to 29 to 53. 
\end{itemize}

We mark the location of the IP with a red square (Fig.~\ref{fig:nikkodown}). The rest of the images (Fig.~\ref{fig:nikkodown}) are reflections off the outer wall of the Beam Pipe. The red ellipses correspond to the waist of the Beam Pipe
at the IP.

\begin{figure}[ht]
    \centering
    \includegraphics[width=0.8\textwidth]{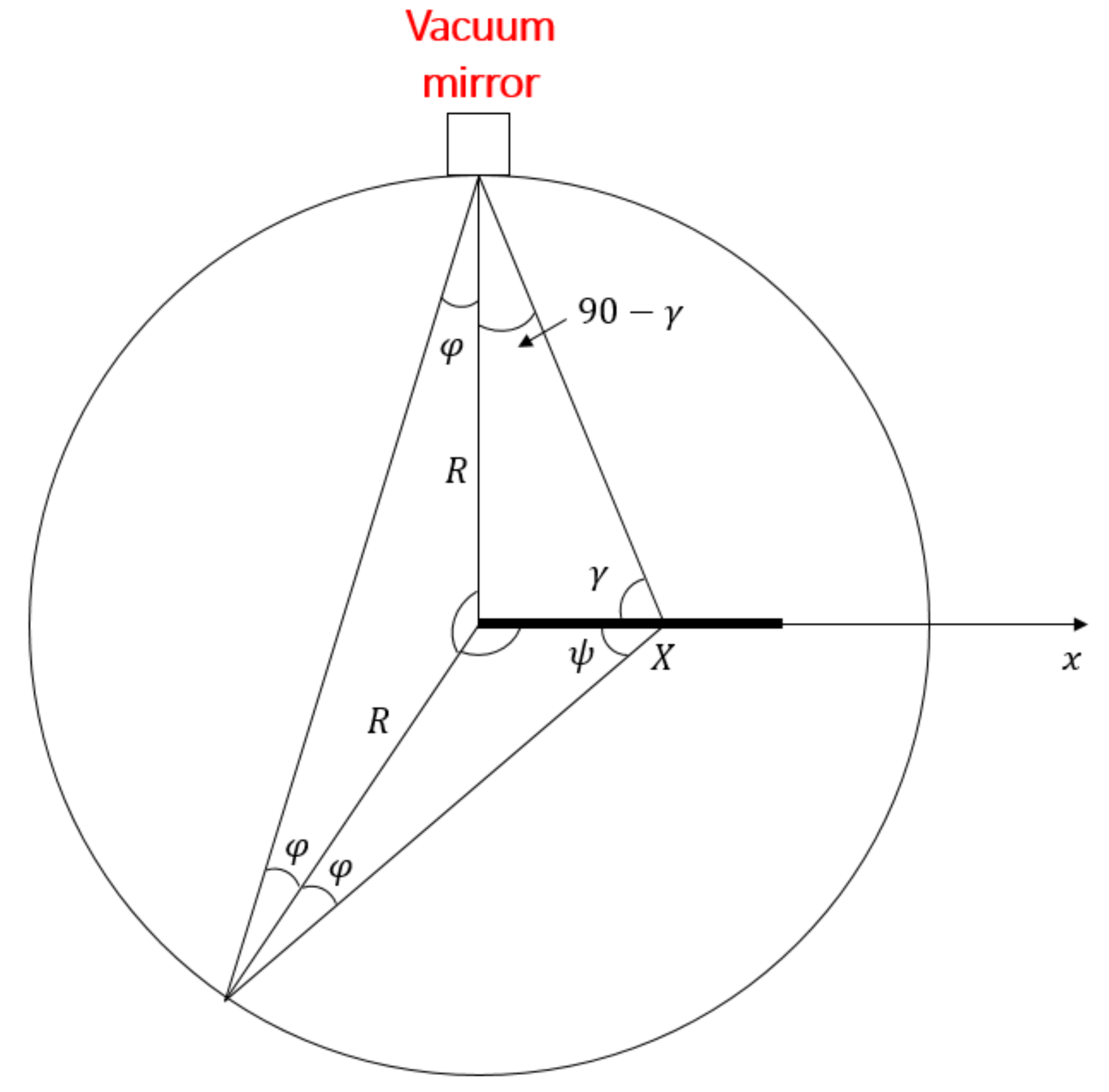}
    \caption{Transverse geometry of synchrotron radiation reflections in the
      Beam Pipe. The position of the radiating point along the sweep
      is defined by $X$, and the angles $\phi$ and $\psi$ are defined.}
    \label{fig:refl}
\end{figure}

We calculate which angles allow for reflection into our lower telescope window, starting from the nominal beam located at $(-x,0)$ as it sweeps through the last dipole. We find that the equation relating the reflection angles $\psi$ on the Beam Pipe reads ($r$ is the inner radius of the Beam Pipe) (see Fig.~\ref{fig:refl})

\[\frac{x}{\sin{\phi}}=\frac{r}{\cos{(3\phi)}},\;\; \psi=\pi/4-2\phi.\]

This equation can be tranformed into a cubic equation in $\cos^2{\phi}$. One or three solutions are possible.
% xxx
By varying $x$ to evaluate the properties of the reflection sweeps, we find that for $x=0$ (the last point on the sweep) the one solution is $\psi=\pi/2$, indicating a reflection off the top or bottom of the Pipe. These reflections can only be seen very faintly above and below the main sweep. They are faint
because the radiation angle is triple the angle of the main sweep, and they are also slightly to the left of the main sweep because the reflection from the early part of the sweep will be slightly towards the right.

With increasing $x$, two solutions become available. The value of $\psi$ becomes $\pi/4, 5\pi/4$ for $x=1$. Therefore reflections of sweeps that cover an interval of $x$ will have a slope. We can clearly see two such reflections in Fig.~\ref{fig:nikkodown}, Reflections III and IV.

Reflection V can not be reconciled with any expected reflection from the last magnet. We believe it is a triple reflection from the second to last magnet. We calculate that such a reflection would show up in $\theta_y$ approximately 1.5 mrad above the true sweep. The observed sweep is 1.0 mrad above the true sweep.

\subsection{Oho Up}

The Oho data, observing the HER, or electron, beam, are similar for both the Up and Down detectors. The last bend in this line is a soft bend, so that the sweep is only 30\% wider than our angular resolution. The beam in both telescopes will present itself as a small, very bright spot (because all the light from the bend is concentrated in two scan points). The straighter, more reflective Beam Pipe creates more diffuse reflections. 

\begin{figure}[ht]
\centering
\includegraphics[width=0.8\textwidth]{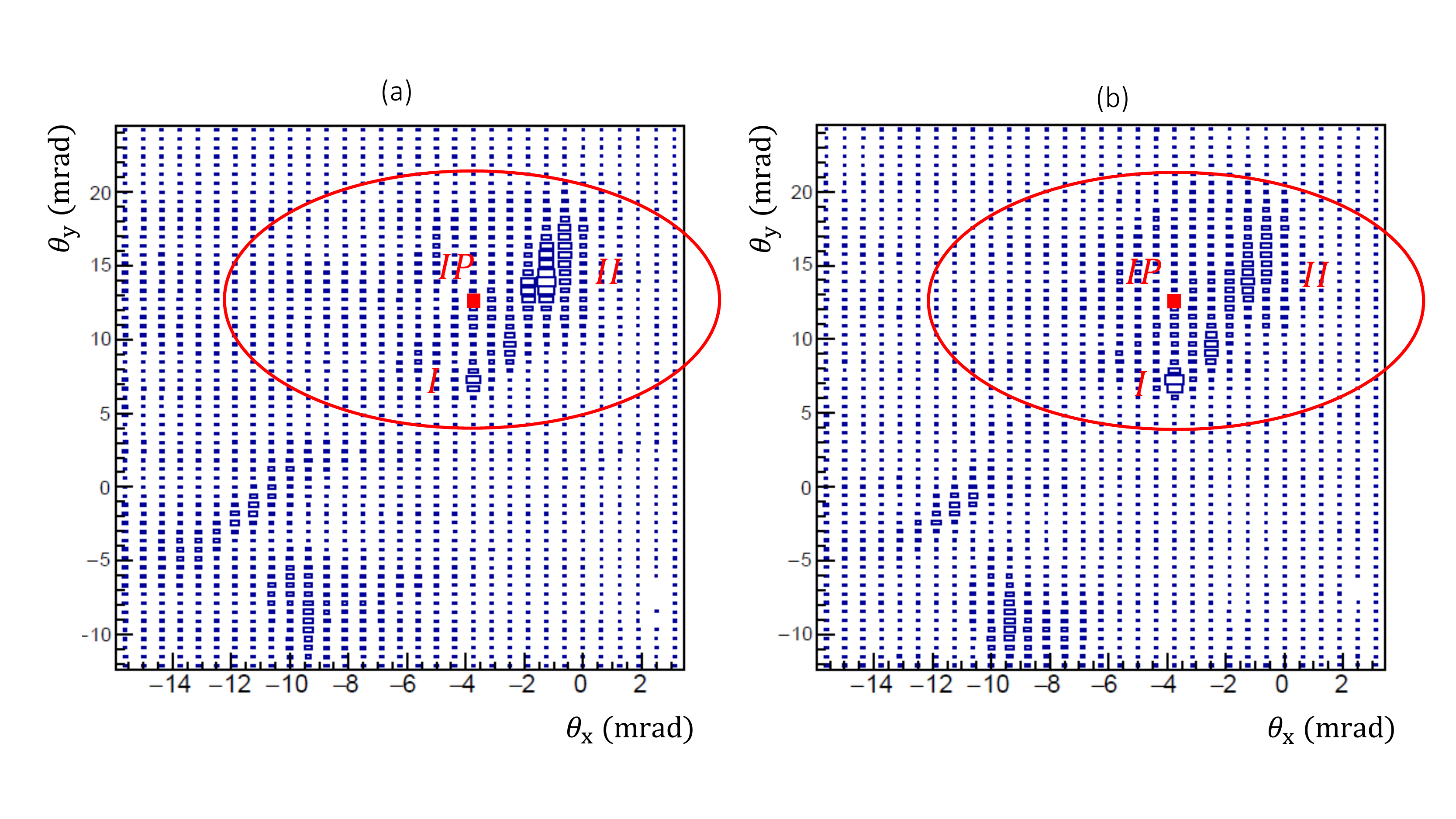}
\caption{Oho Up angular scan with ``UV'' PMT. The IP is marked as a
red square.}
\label{fig:ohoup}

\end{figure}

The Oho Up data including the IP are shown in Fig.~\ref{fig:ohoup}. Here, too, broad reflections are seen. The one spot consistent with the sweep is clear. Oho Up showed widespread PMT saturation, so that this image was taken with the ``UV'' PMT, which only accepted less than 1\% of the total light. To obtain cross comparisons with Nikko Down, we used data from 20 mA beams, which had only one PMT (``green'') showing saturation. Oho Up is the only telescope where the $(x,y)$ mapping is inverted, and it shows it by being the only one with a strong $y-$ polarization at the Box.

\subsection{ Nikko Up and Oho Down.}
In the case of Nikko Up, we were unable to obtain images consistent with an
angular scan. This was due to a mechanical problem with all the Primary mirrors,
which were built from scratch out of metal at Wayne State to make sure
they would withstand the high radiation environment. In a commercial remotely
controlled mirror, the mirror support is made of somewhat flexible plastics
which yields and can transform the linear motion of the stepper motors
into the spherical motion of the rotating mirror. In our case, the first
version of the mechanism involved one square brass rod sliding into a slightly
larger rod, and this transmission was sensitive to slight misalignments between
stepper motors and mirrors, to the point where one of the four
primary mirrors (the Nikko Up) would move irregularly.
The current version, where the transmission is made
with sliding forks operating the mirror support knobs, is extremely robust.

Nikko Up is also the only one of the four telescopes where the remotely
controlled secondary mirror is 100 cm from the primary (every other
secondary is within 30 cm). With the primary stuck in an uncertain direction,
the secondary mirror here
is too far from the vacuum mirror to be usable for the angular scan.
Having no back up scanning, we obtained images with bands which were consistent
with one of the two mirror controls not actually moving.

Oho Down produced good images, but with rates which were 30 times lower than
Oho Up. Diode calibration data, which had been used to calibrate the PMTs,
were re-analyzed to discover that diode data, too, were weaker by a factor of 30
with respect to Oho Up. This discovery shows that the attenuation takes place
in the Optical Channel, and not inside the Beam Pipe where it could not be
fixed. To solve this problem, we have established a new protocol for alignment
were the diode intensity is measured at every step of the Optical Channel
assembly, so that intensity losses are immediately detected and remedied. The
detector as presented here had only final diode intensity measurements.

\section{Large angle synchrotron radiation.}
In order to discuss quantitatively the obtained images,
it is important to consider some properties of the large angle SR, which we review in this section. Ref.~\cite{bonvicbn} discusses the properties of large angle synchrotron radiation. It considers the classical formulae from Ref.~\cite{jackson}, which only offer the intensity as a function of the elevation angle $\psi$ above the sweep. In Gaussian units, the approximation $\gamma\theta\gg 1$ and the fractional Bessel functions in the limit of large argument, one electron will radiate according to

\begin{equation} \label{eq:classicalAsymptoticSR}
\frac{d^{2}U}{d\omega d\psi} \sim \frac{e^2}{\pi c} \bigg( \frac{\omega \rho \psi}{c} \bigg) exp\bigg( \frac{-2\omega \rho \psi^{3}}{3c} \bigg)
\end{equation}

At large angle, these formulae are unpolarized. 
At the location where the LABM vacuum mirrors are located, which is directly
above or below the beam axis, the elevation angle and the common spherical
angle $\theta$ coincide, and so the latter symbol is used.

Following Ref.~\cite{bonvicbn}, the magnet length $L$ is introduced, the total radiated energy $U=PL/c$, and the differentials are such that $d\psi=4\pi d\theta$.
Making the large-angle substitutions

\begin{equation} \label{eq:SRchangeVariables}
(1 + \gamma^{2} \theta^{2}) \sim u^{2} \,\,\,;\,\,\, \omega \sim \frac{3ct}{\rho \theta^{3}}
\end{equation}

the final double-differential distribution in $(t,u)$ is obtained,

\begin{equation} \label{eq:AsymptoticSRchangeVariables}
\frac{d^{2}U}{du dt} \sim \frac{54 \pi U \rho}{L} \bigg( \frac{t}{u^5} \bigg) exp(-2t)
\end{equation}

The short magnet approximation considers angles much larger than $1/\gamma$, Ref.~\cite{coisso}. In the large angle approximation, the radiation is polarized according to an azimuthal pattern. Using the same notation, the general form for short magnet radiation is

\begin{equation} \label{eq:shortMagnetSRx}
\frac{d^{3}U_{x}}{du dt d\phi} \sim \frac{48 U \rho}{\pi L \gamma} \frac{cos^{2}(2\phi)}{u^{4} t^{2}} 
\end{equation}

and

\begin{equation} \label{eq:shortMagnetSRy}
\frac{d^{3}U_{y}}{du dt d\phi} \sim \frac{48 U \rho}{\pi L \gamma} \frac{sin^{2}(2\phi)}{u^{4} t^{2}} 
\end{equation}

where the angle $\phi$ is taken from the direction of the force
applied to the electron. The full equations also have exponential factors that
suppress radiation at angles larger than the few milliradians considered here.

The transition angle between the two regimes is at around $20/\gamma$ \cite{bonvicbn}. In Table~\ref{tab:geometry}, we see that the angles at which the last dipoles are observed are in the ``short magnet'' region, at 33 and 44 times $1/\gamma$ respectively. We do not have a precise caluclation, but use the formulae above to extract useful comparisons.
From the different angles of observation, we expect a redder, more polarized
spectrum on the positron side, and a less polarized, bluer spectrum on the electron side.

\section{Cross checks.}
We use Eq.~\ref{eq:shortMagnetSRx}, integrated over
the visible spectrum and the solid angle subtended by the primary mirrors,
and divided over the number of pixels in a sweep,
to calculate SR rates $(Hz/mA)$ on each side and compare
them against the true rates calculated according to
Eq.~(3). Only the rates in pixels identified as belonging to a sweep are used.

To compare the experimental rates
with the calculations, we use the LABM efficiency calculations from
Ryan Gillard's Thesis~\cite{ryan}, which finds that the overall photon
detection efficiency varies from 2 to 7\% depending on wavelength,
averaging about 4\%.
We had to re-run the Monte Carlo because we made the operative decision to use (in Phase I alone) wider ruled gratings than previously planned, and in the process we sacrificed one PMT for background measurements. The overall detector efficiency does not change, but the visible photons are currently detected by three out of four PMTs in each array. In Phase II, having found that backgrounds localized in the Optics Box are negligible, we will return to the nominal gratings.

The Up/Down comparisons would be very useful to extract information about the relative angle of radiation with respect to the detector but,
as stated above, one detector on each side had problems.

Our results are summarized in Table~\ref{tab:cross} and show
good agreement (to about 20\%, consistent with the error in determining
the exact number of pixels in a sweep) for
the total rates on both sides. Also the total rates
are very similar on the positron and electron side, both as measured
and as calculated. This is consistent with both telescopes
observing the true beam, and with them having similar efficiency.

The ratio green/blue $R_{23}$ is also given (the ratio of true rates of the second
and third PMT), and also the polarization ratio, which is the $R_x/R_y$ ratio
for a given color.
We find that the positron telescope does see highly polarized light consistent
with Eqs.~(7-8) and that the ratio increases rapidly with increasing frequency.
However, the electron telescope sees a bluer light with little polarization,
consistent with a smaller angle of observation but not consistent with the
azimuthal pattern expected in large anlge radiation. While this is
qualitatively correct, we do not have a model to explain in detail the large
polarization difference between the two sides. It is noted that in any case,
backgrounds will be measured directly during Phase II, during one beam runs,
so that a detailed model of large angle SR radiation is not needed.

\begin{table}%{htb}
\centering
  \caption{Comparison of rates and calculated intensities. The quantities are described in the text.}
\label{tab:cross}
%\centering 
\begin{tabular}[t]{|c|c|c|}
\hline
 Quantity & Nikko Down & Oho Up  \\
 \hline

Exp. total Rate (Hz/mA) & 9E4  & 1.4E5    \\
Theor. total Rate (Hz/mA) & 1.0E5  & 1.3E5  \\
$R_{23x}$ &2.1 & 1.1  \\
$R_{23y}$ &4.0  & 0.9 \\
$R_x/R_y$(red) &2.0&1.5 \\
$R_x/R_y$(green) &3.3&1.6 \\
$R_x/R_y$(blue) &7.3&1.4 \\

\hline
\end{tabular}
\end{table}

\section{Signal expectations.}
In this Section we compare the rates at the calculated IP positions against
future beamstrahlung signals. Onyl the normalized rate of the pixel closest
to the calculated IP is considered.

Beam parameters are listed in Table~\ref{tab:beams} for June 2016. We have assumed 1A currents for each beam. The total beamstrahlung yield, the fractional energy acceptance of the LABM, and the calculated photoelectron rates (summed over all PMTs) are also reported. We have assumed 1A currents for each beam. The total beamstrahlung yield, the fractional energy acceptance of the LABM, and the calculated photoelectron rates (summed over all PMTs) are also reported.

\begin{table}%{htb}
\centering
  \caption{Beam parameters and LABM yields for Phase I and Phase II.
LABM $\epsilon$ means LABM energy acceptance, the fraction of total radiated 
energy intercepted by the vacuum mirror. LABM rates refer to the total
photoelectron rates(summed over all phototubes), while LABM S/B and
compare the currently
measured background rates to the expected signal rates. HER, LER I represent
  Phase I parameters, HER, LER II Phase II parameters.}
\label{tab:beams}
%\centering 
\begin{tabular}[t]{c|c|c|c|c}
\hline
 Quantity & HER I & LER I & HER II & LER II \\
 \hline
Current(A) & 1 & 1 & 2.6 & 3.6  \\
Frequency (MHz) & 250 & 250 & 250 & 250 \\
Bunch pop(10$^{10}$) & 2.5 & 2.5 & 6.5 & 9 \\
$\sigma_x$($\mu$m) & 430 & 500 & 11 & 10 \\
$\sigma_y$($\mu$m) & 36 & 112 & 0.062 & 0.048 \\
$\sigma_z$(mm)             & 5 & 6 & 5 & 6 \\
LABM rates (Hz) & (-) & ( - ) &(1.3E8,6.5E7) & (3.3E8,1.1E8) \\
Background(Hz)  & (1E7,5E6) & (8E5,1E5) & (2.6E8,1.3E8) & (2E7,2.5E6) \\
LABM S/B & - & - &(0.5,0.25) & (16, 44) \\
\hline
\end{tabular}
\end{table}

We note the S/B greater than 10 for Phase II on the LER side and of order 1
on the HER side. Regrettably the HER side has a significant amounts of
reflection right where the IP is.

Further, we are extrapolating the present rates being extrapolated to Phase II. However, there are currently no SR masks installed in SuperKEKB. It is expected that the masks will strongly reduce the noise for rays coming from an angle consistent with the IP. But even without masks, the LABM would be able to perform a variety of studies with background subtraction techniques. It is also noted that
the Phase I measurements do not include the expected Touschek halo accompanying
the beam, which may contribute extra backgrounds when crossing the final
quadrupoles off-axis.

\section{Conclusions.}
The LABM has shown that two of the telescopes work close to specifications. The detector is dark and has worked continuously for 4 months.
The problems encountered with the other two telescopes have been overcome by substituting
some mechanical parts and by changing the alignment protocol.

The main results are:
\begin{itemize}
  
\item the direct measurement of expected radiation patterns through
  angular scans. This shows first
  that the whole system works with some redundancy (scans were performed
  by moving either the primary or secondary mirrors), and that the
  modeling of radiation inside the Beam Pipe is of sufficient quality
  for both the positron and the electron side.
\item from the patterns, the geometric determination of the IP, which is crucial
  for the LABM.
\item from the determination of the IP, the direct measurement of the
  expected Phase II backgrounds,
  although Phase II beams will also have a smaller, spectrally different
  contribution from the Touschek halo, absent in Phase I, which may contribute
  some extra background from radiation in the Final Focus quadrupoles. Under the
  assumption that the Touschek halo is not dominant, the backgrounds are
  sufficiently low that LABM operations will be possible. They will be further
  reduced by a planned change in collimators.
\end{itemize}

\section*{Acknowledgment}

We thank Y. Funakoshi for useful discussions.
This project has received support from the US-Japan Science and Technology
Cooperation Program, and grants
Fronteras de la Ciencia Conacyt FOINS-296,
Ciencia Basica Conacyt  CB-180023, 
PROFAPI  UAS   2013-144, and University of Tabuk grant 1436/052/2.

\section*{References}

\bibliography{gbbibfile}

\end{document}